# Linear Dynamic Polarizability and the Absorption Spectrum of an Exciton in an Aharonov-Bohm Quantum Ring

A.V. Ghazaryan[1*], A.P. Djotyan[1], K. Moulopoulos[2] and A.A. Kirakosyan[1]

[1] *Department of Physics, Yerevan State University, 1 Al. Manoogian, 0025 Yerevan, Armenia*
[2] *Department of Physics, University of Cyprus, PO Box 20537, 1678 Nicosia, Cyprus*

*a.ghazaryan@ysu.am

Recent developments in the fabrication of semiconductor nanostructures have led to the construction of III–V semiconductor volcano-shaped nanorings [1]. Quantum correlations in such systems with interacting charged particles moving in multiply-connected spaces, and especially in the presence of magnetic Aharonov–Bohm (AB) fluxes, make these systems representative of an important topic in Condensed Matter Physics and of a new area of research that is potentially useful for applications.

Furthermore, the optical properties of confined electrons and holes in nanostructures, particularly in semiconductor quantum rings (QRs) in an external magnetic field, have been a subject of increased interest in recent years [2-5]. Besides focusing on the fundamental energetics of quasiparticles confined in QRs, investigations of the optical properties of QRs have great potential for creating new functional devices. In particular, the theoretical study of electric polarizability and the light absorption spectrum of an exciton system is especially helpful for the development of experiments, as it provides the opportunity to compare with experimental measurements and directly verify the validity of the peculiarities of the energy levels revealed by theoretical calculations for such a multiply-connected system.

In this work we theoretically solve the problem of an exciton (with particles interacting by a delta potential) in a one-dimensional QR in the presence of an AB flux. For solving this problem we first bring its Schrödinger equation into a decoupled form by using the definition of the center of mass and relative variables. The equation connected with the center of mass variable represents a free particle and has a trivial solution. In order to solve the equation connected with relative variable we use a more straightforward method compared to earlier studies [3-5], which involves a procedure of determining the wave function of the system at the points where the potential is zero, and then of imposing boundary conditions at the singular point connected with the delta function. By this we find the energy spectrum and the associated eigenstates together with other physical properties of the system in closed analytical forms. In particular, the expression that determines the energy spectrum for relative motion has the following form

$$\cos\left(2\pi(f-l)\right) - \cos\left(2\pi\sqrt{B}\right) = \frac{U}{2\Delta\sqrt{B}}\sin\left(2\pi\sqrt{B}\right), \tag{1}$$

where

$$B = \frac{E_r}{\Delta} = \frac{E - E_c}{\Delta}, \quad f = \frac{Q_r \Phi}{hc}, \quad \Delta = \frac{\hbar^2}{2\mu_r R^2}, \tag{2}$$

$$l = \frac{m_2}{m_1 + m_2}n_1 - \frac{m_1}{m_1 + m_2}n_2, \quad \mu_r = \frac{m_1 m_2}{m_1 + m_2}, \quad Q_r = \frac{q_1 m_2 - q_2 m_1}{m_1 + m_2}, \tag{3}$$

and $m_1(m_2)$ and $q_1(q_2)$ are the first (second) particle mass and charge respectively, $n_1$ and $n_2$ are integers, $U$ is the interaction potential constant, $\Phi$ is the magnetic flux threading the ring and $E = E_r + E_c$ is the total energy, with $E_r$ being the energy of the relative motion and $E_c$ the center of mass energy.

In Fig. 1 the dependence of the excited states energy levels on flux parameter $f$ for different values of the interaction potential constant $U$ and for total orbital angular momentum $N = 0$ are shown. As can be seen, for values of $f = 0.5$ for the first (and in the case $f = 0$ for the second) excited state, the energy levels do not depend on the value of the interaction potential and they have constant values. These states are formed when the difference $\cos(2\pi(f-l)) - \cos(2\pi\sqrt{B})$ and $\sin(2\pi\sqrt{B})$ in Eq. (1) become zero simultaneously, so that Eq. (1) is satisfied regardless of the value of the interaction potential. This condition is satisfied when $f - l = n/2$, where $n$ is an arbitrary integer, and for these values of $f$ the solution of (1) turns out to be $B = (2m+n)^2/4$, where $m$ is another integer. These are essentially free particle energies (for the relative system) and their existence originates from the particular form of the wavefunctions whenever the above condition holds: it is shown that for these energy values the wavefunction becomes zero at relative angular variable $\varphi = 0$ (namely, when the particles are in contact). This means that, for these special energies, the "relative particle" does not have any contact with the potential (because of the delta potential form of the interaction) and is therefore completely free.

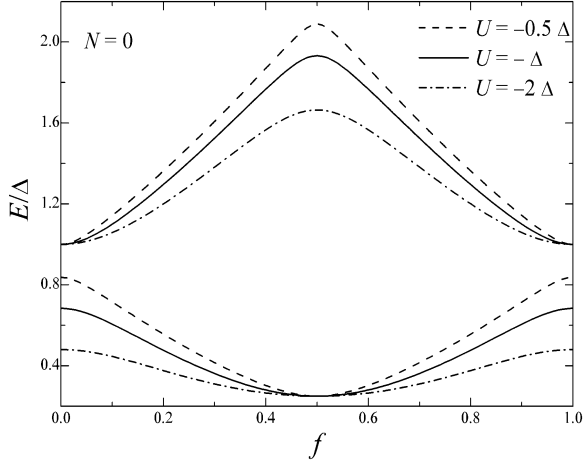 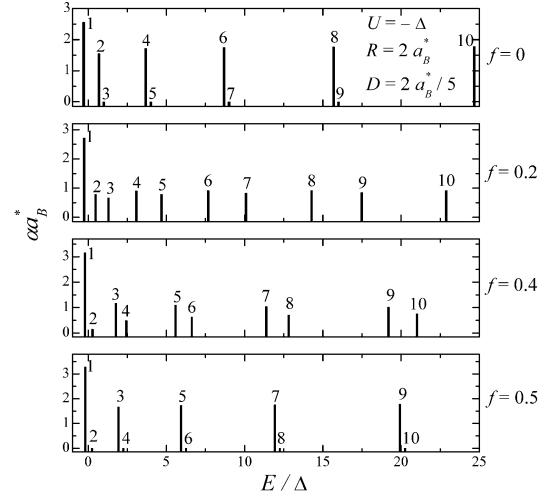

**Figure 1.** The dependence of the excited states energies on flux parameter $f$ for different values of the parameter $U$ and for pair-angular momentum $N = 0$.

**Figure 2.** The values of the excitonic absorption coefficient $\alpha$ in units of inverse effective Bohr radius in the QR with the radius $R = 2a_B^*$ and the thickness $D = 2a_B^*/5$ for several states and for different values of the flux parameter $f$.

After finding the energy spectra of the exciton in this system, we then calculate the dynamic linear electric polarizability and the absorption coefficients, which we believe can be very helpful for experimentalists for checking the validity of the theory developed in this paper, and also for comparing with previous investigations [3-5].

In the calculation of optical properties, we assume that the ring has finite thickness and a square cross-section with linear size $D < a_B^*$ ($a_B^* = \hbar^2\varepsilon/m_e e^2$ is the effective Bohr radius in bulk semiconductor, with $\varepsilon$ being the static dielectric constant of the bulk material). Due to the small thickness (in comparison to the radius of the ring and to the characteristic length of bound states) we assume that the interaction between particles is one-dimensional and takes place only along the perimeter of the ring (and we also assume that the confining potential creates appropriate infinite barriers for the electron and the hole on the ring surface). As a system for our consideration we take the quantum ring made of GaAs and we use the corresponding values of the parameters for this material $m_e = 0.067 m_0$, $m_{lh} = 0.087 m_0$, $E_g = 1.424 eV$, $\varepsilon = 13.18$, $P_L^2 = |P_{cv}|^2/m_0 = 14.4 eV$ [6]. For the radius and thickness of the ring we take values that are comparable to those observed in experiment [7], where Aharonov-Bohm oscillations of a neutral exciton in a single self-assembled nanoring were experimentally observed, namely $R = 2a_B^*$ and $D = 2a_B^*/5$.

In Fig. 2 the values of the excitonic absorption coefficient in the QR with radius $R = 2a_B^*$ and linear size $D = 2a_B^*/5$ for several states are shown and for different values of the flux parameter $f$. As can be seen, when the magnetic flux parameter is equal to $f = 0$ or $f = 0.5$ some states have zero absorption coefficients and we have shown that this is closely related to the peculiar behavior of excited energy levels described above. From Fig. 2 we also observe that the magnetic flux changes the values of absorption coefficients dramatically and that by changing the value of magnetic flux parameter from 0 to 0.5 dark exciton states transform into bright ones and vice versa.

Although this feature is the special property of the delta potential interaction, we believe that to some extent it can also be observed for a general short range interaction, something that can be tested experimentally.